\documentclass[prd,superscriptaddress,altaffilletter,showpacs,nofootinbib]{revtex4}
\usepackage{amsfonts}
\usepackage[dvips]{graphicx}
\usepackage{amsmath}

\begin{document}

\title{Soliton structures in a molecular chain model with saturation}
\author{M. Aguero}
\email{mag@uaemex.mx}
\affiliation{Facultad de Ciencias, Universidad Autonoma del Estado de Mexico, Instituto
Literario 100 Toluca, Edo.Mex. 50000, Mexico}
\author{R. Garc\'{\i}a-Salcedo}
\email{rigarcias@ipn.mx}
\affiliation{Centro de Investigacion en Ciencia Aplicada y Tecnologia Avanzada - Legaria
del IPN, Av. Legaria 694, Col. Irrigacion, 11500, D.F., M\'{e}xico.}
\author{J. Socorro}
\email{socorro@fisica.ugto.mx}
\affiliation{Facultad de Ciencias, Universidad Autonoma del Estado de Mexico, Instituto
Literario 100 Toluca, Edo.Mex. 50000 and Instituto de F\'{\i}sica de la
Universidad de Guanajuato, A.P. E-143, C.P. 37150, Le\'on, Guanajuato,
M\'exico.}
\author{E. Villagran}
\email{edgarvillagran@hotmail.com}
\affiliation{Facultad de Medicina y de Ciencias, Universidad Autonoma del Estado de
Mexico, Instituto Literario 100 Toluca, Edo.Mex. 50000, Mexico}

\begin{abstract}
In the present work, we study, by means of a one-dimensional lattice model,
the collective excitations corresponding to intra molecular ones of a chain
like proteins. It is shown that such excitations are described by the
Nonlinear Schrodinger equation with saturation. The solutions obtained here
are the bell solitons, bubbles, kinks and crowdons. Since they belong to
different sectors on the parametric space, the bubble condensation could
give place to some important changes of face in this kind of nonlinear
system. Additionally, it is shown that the limiting velocity of the solitons
is the velocity of sound waves corresponding to longitudinal vibrations of
molecules.
\end{abstract}

\maketitle

\section{Introduction}

In the present days the research to the dynamic properties of one
dimensional molecular chains have increased. The structure of a great amount
of macromolecules is represented by subunits with mutually weak, slightly
flexible bounds connecting them with each other. The important biological
structures of this type are RNA, proteins and DNA polymer chains. The
peculiarity of bio-polymers is that, they are heterogeneous, and their
elementary subunits have complex structures and carry long-lived nonlinear
excitations. As it is well known the propagation of energy and electrons in
protein molecules is the crucial factor for maintaining the life of
biological systems. So, the problem of storage and transportation of energy
through protein chains arises. The energy used in biological cell comes from
the energy of liberation during the process of hydrolysis of
adenosinetriphosphate (ATP) molecular structures. The energy of this process
is of approximately 0,42 eV (or 2250 cm$^{-1}$). Proteins consist of chains
of hydrogen-bonded peptide groups, three of these chains in a helical
arrangement define the $\alpha -$helix structure \cite{ev}. In his seminal
work, Davydov proposed an explanation of the fundamental transportation
problem of energy released by hydrolysis of adenosinetriphosphate and
transferred to proteins in biological systems \cite{davidov}. This energy
remains localized and moves along the protein chains at a reasonable rate to
perform useful biological functions. It could be trapped and transported in
proteins as quanta of the intra molecular C=O stretching mode, the so called
amide-I vibration, with excitation energy around 1650cm$^{-1}$. The
localized spatial region where the energy is trapped can propagate along the
protein chain, in such a way that a soliton-like mechanism for energy
transport is possible. This problem of transporting energy from one point to
another inside the cell is a long-standing problem that remains of great
interest.

Besides, from experimental point of view, we can also discover a lot of
contributions that are directly related to a similar phenomenon in the DNA.
For instance, in the experimental work on short DNA\ rings e.g., Refs. \cite%
{ring, ring2} the kink tendency of DNA\ sequences were studied. In the last
years the great amount of works devoted to the nonlinear dynamics of DNA\
shows that this area is an active field, for example theoretical proposals
of wrapping DNA around the nucleosome, where kinks play a great rule, were
proposed in various papers, see the Refs.\cite{agrawal,cuevas,chris}.
Concerning the $\alpha -$helical protein dynamics, some works have been
dedicated to study this system including high order excitations and
different molecular interactions, in the discrete and continuum level, \cite%
{daniel,scott}.

In this contribution we investigate soliton-like structures within the
framework of a certain generalization of the Davydov's model, considering
the case of neighboring interactions as the same class before and after a
peptide group. In the next section we briefly expose the modified Davydov's
model leading to the Hamiltonian that will be used in the later. Section III
is devoted to derive the nonlinear cubic - quintic Schrodinger equation by a
suitable transformation from the original nonlinear Schrodinger equation
with saturation. The soliton structures of this equation with some specific
characteristics are presented in section IV. Finally, in the last section we
deliver some comments.

\section{Davydov's Model.}

Due to its transparency and seminal properties, Davydov`s model continues to
encourage intense work regarding the research of the nonlinear treatment of
molecular systems \cite{davidov}. In his pioneer works he and co-workers
demonstrated that the corresponding nonlinear equations for the molecular
excitation in the quantum treatment admit solitonic structures. Indeed they
assumed that the energy transportation in proteins is carried out by means
of transportation of amide - I vibration. In what follows we consider an
infinite chain of weakly bound molecules (or groups) with a mass $M$ and a
distance $R$ from each other. Internal excitations of molecules (electronic
or vibrational) are characterized by an energy and an electric dipole moment
$d$ directed along the chain. The internal excitation of molecules and their
motion around equilibrium positions are inseparably linked.

In the case of a one-dimensional chain, only interactions between neighbor
molecules are taken into consideration. Below, we follow the modeling done
by Davydov and his co-workers. If the intramolecular excitation has the
energy $\varepsilon $, the collective excitations of this model can be
described by the Davydov's Hamiltonian:

\begin{eqnarray}
H&=&\sum_{n}\left[ (\varepsilon -D_{n})B_{n}^{+}B_{n}-J\left(
B_{n+1}^{+}B_{n}+B_{n+1}B_{n}^{+}\right) \right]  \notag \\
& & +T+U,  \label{hami}
\end{eqnarray}
where index $n$ labels the molecule that occupies position $r_{n}$ in the
chain, while $B_{n}^{+}$ y $B_{n}$ are creation and annihilation of
intramolecular excitation boson operators. The quantity $J=2d^{2}R^{-3}$
characterizes the transition of intramolecular excitation due to the
resonant interactions while $d$ is the electric dipolar moment. The last two
terms in (\ref{hami}), as usual, correspond to the kinetic and potential
energies of the longitudinal displacements.

When the $i$-th molecule is excited, the static interaction with neighboring
molecules of this molecule changes. This is reflected by the introduction of
the function $D_{n}$. The displacement $\rho _{n}$ from the equilibrium
distance $R$ in the state $\mid \Psi \rangle$ is defined by the expression

\begin{equation}
\rho _{n}=R-(r_{n}-r_{n-1}).  \label{displa}
\end{equation}%
In the state $\mid 0\rangle \;$ without intramolecular excitation, the chain
has periodicity and the intermolecular distances are $R $.

Let us now consider the function $D_{n}$, which in the nearest neighbors
interaction limit has the following form

\begin{eqnarray}
D_{n} &=&\mathfrak{D}_{n}\left( \left\vert r_{n-1}-r_{n}\right\vert \right) +%
\mathfrak{D}_{n}\left( \left\vert r_{n}-r_{n+1}\right\vert \right)  \notag \\
&\approx &\left( 1+\frac{\beta }{R}\rho _{n}+\frac{\beta \gamma }{2R^{2}}%
\rho _{n}^{2}\right) D,  \label{displace}
\end{eqnarray}%
where $D\equiv 2\mathfrak{D}_{n}\left( R\right) ,\beta ,\gamma $ are
parameters of the theory.

The potential energy of the molecules in the non excited chain is chosen in
a harmonic approximation under the assumption that the constant spring $%
\omega \ $is the same for all of them. In this case we can express the
potential energy as

\begin{equation*}
U=\frac{1}{2}\omega \sum\limits_{n}\rho _{n}^{2},
\end{equation*}%
and the kinetic energy can be written as

\begin{equation*}
T=\frac{1}{2}M\sum\limits_{n}\left( \dot{r}_{n}\right) ^{2}=\frac{1}{2}%
M\sum\limits_{n}\left( \sum\limits_{-\infty >l\leq n}\dot{\rho}\right) ^{2},
\end{equation*}%
where dot over the letter represents temporal derivative, $\dot{\rho}\equiv
\frac{d\rho }{dt}.$ According with the quantum mechanics treatment the
collective interactions of interest can be described by the wave function

\begin{equation*}
\left\vert \Psi \right\rangle =\sum\limits_{n}\psi _{n}\left( t\right)
B_{n}^{+}\left\vert 0\right\rangle ,
\end{equation*}%
where coefficients $\psi _{n}\left( t\right) $ are normalized as $%
\sum\limits_{n}\left\vert \psi _{n}\left( t\right) \right\vert ^{2}=1.$
These coefficients characterize the distribution of excitations along the
molecular chain, $\psi _{n}$ being the probability of finding the quantum
system in the site $n$ or $\left\vert \psi _{n}\left( t\right) \right\vert
^{2}$ being the density of probability of finding the excitation. The
equation for determining these wave functions can be obtained from the Schr%
\"{o}dinger equation

\begin{equation*}
i\hbar \frac{\partial }{\partial t}\left\vert \Psi \right\rangle
=H\left\vert \Psi \right\rangle ,
\end{equation*}%
that can be reduced, using the explicit form of the operator $H$, Eq. (\ref%
{hami}), the fact that the functions $B_{n}^{+}\left\vert 0\right\rangle $
correspond to different values of $n$ and are orthogonal each other, to
obtain the following system of equations

\begin{eqnarray}
i\hbar \frac{\partial \psi _{n}}{\partial t} &=&\left[ \varepsilon
+T+U-\left( 1+\frac{\beta }{R}\rho _{n}+\frac{\beta \gamma }{2R^{2}}\rho
_{n}^{2}\right) D\right] \psi _{n}  \notag \\
&&-J\left( \psi _{n+1}+\psi _{n+1}\right) .  \label{aa}
\end{eqnarray}%
The functional \textit{i.e.} the hamiltonian that can be associated with
this equation of motion can be written as $F=\left\langle \Psi \right\vert
H\left\vert \Psi \right\rangle $

\begin{eqnarray*}
F &=&\sum\limits_{n}\left\{ \left[ \varepsilon +T+U-\left( 1+\frac{\beta
\rho _{n}}{R}+\frac{\beta \gamma \rho _{n}^{2}}{2R^{2}}\right) D\right] \psi
_{n}^{\ast }\psi _{n}\right. \\
&&\left. -J\psi _{n}^{\ast }\left( \psi _{n+1}-\psi _{n-1}\right) \right\} .
\end{eqnarray*}%
Following Toda \cite{Toda}, it is convenient to associate the displacements $%
\rho _{n}$ with their canonically conjugate variables $s_{n}=\frac{\partial T%
}{\partial \dot{\rho}_{n}}=-M\sum\limits_{l\geq n}\dot{r}_{l}$. Then the
kinetic energy can be expressed in terms of these new variables as

\begin{equation*}
T=\frac{1}{2M}\sum\limits_{n}\left( s_{n}-s_{n-1}\right) ^{2}.
\end{equation*}

In the next section we will see the derivation for the nonlinear
Schr\"odinger equation (NSE) with a saturation term.

\section{NSE equation with saturation}

Now, we can derive the equation of motion for the displacements and for
their canonical conjugate variables $s_{n}$, for this we consider that

\begin{eqnarray}
\dot{\rho}_{n} &=&\frac{\partial F}{\partial s_{n}}=\frac{1}{M}\left(
2s_{n}-s_{n+1}-s_{n-1}\right) ,  \notag \\
\dot{s}_{n} &=&-\frac{\partial F}{\partial \rho _{n}}=-\omega \rho _{n}+%
\frac{\beta D}{R}\left\vert \psi _{n}\left( t\right) \right\vert ^{2}+\frac{%
\beta \gamma D}{R^{2}}\rho _{n}\left\vert \psi _{n}\left( t\right)
\right\vert ^{2}.  \label{dis2}
\end{eqnarray}%
After eliminating the variables $s_{n}$ from the preceding system of
equations, we find the equation for the displacement

\begin{eqnarray}
\ddot{\rho}_{n} &=&-\frac{\omega }{M}\left( 2\rho _{n}-\rho _{n+1}-\rho
_{n-1}\right)  \notag \\
&&+\frac{\beta D}{RM}\left( 2\left\vert \psi _{n}\right\vert ^{2}-\left\vert
\psi _{n+1}\right\vert ^{2}-\left\vert \psi _{n-1}\right\vert ^{2}\right) +
\notag \\
&&+\frac{\beta \gamma D}{R^{2}M}\left( 2\rho _{n}\left\vert \psi
_{n}\right\vert ^{2}-\rho _{n+1}\left\vert \psi _{n+1}\right\vert ^{2}-\rho
_{n-1}\left\vert \psi _{n-1}\right\vert ^{2}\right) .  \label{dis3}
\end{eqnarray}%
The system of equations ( \ref{aa}) and (\ref{dis3}) defines the collective
excitations and deformation of a chain.

For an analytical treatment if we are interested on the distribution of
excitations along the chain, we turn to the analysis in the continuum limit.
For doing this, let us introduce the dimensionless variable $\xi =\frac{r}{R}
$ and the continuous functions as usual $\rho \left( \xi ,t\right) $ and $%
\psi \left( \xi ,t\right) $ such that
\begin{equation*}
\rho \left( n,t\right) =\rho _{n}\left( t\right) ,\;\;\;\;\;\;\;\;\;\;\;\psi
\left( n,t\right) =\psi _{n}\left( t\right) .
\end{equation*}%
Expanding $\rho \left( \xi \pm 1\right) $ and $\psi \left( \xi \pm 1\right) $
in series in the standard manner
\begin{eqnarray*}
\rho \left( \xi \pm 1,t\right) &\approx &\rho \left( \xi \right) \pm \frac{%
\partial \rho \left( \xi ,t\right) }{\partial \xi }+\frac{1}{2}\frac{%
\partial ^{2}\rho \left( \xi ,t\right) }{\partial \xi ^{2}}, \\
\psi \left( \xi \pm 1,t\right) &\approx &\psi \left( \xi \right) \pm \frac{%
\partial \psi \left( \xi ,t\right) }{\partial \xi }+\frac{1}{2}\frac{%
\partial ^{2}\psi \left( \xi ,t\right) }{\partial \xi ^{2}}, \\
\left\vert \psi \left( \xi \pm 1,t\right) \right\vert ^{2} &\approx
&\left\vert \psi \left( \xi ,t\right) \right\vert ^{2}\pm \frac{\partial }{%
\partial \xi }\left\vert \psi \left( \xi ,t\right) \right\vert ^{2}+\frac{1}{%
2}\frac{\partial ^{2}}{\partial \xi ^{2}}\left\vert \psi \left( \xi
,t\right) \right\vert ^{2},
\end{eqnarray*}%
and keeping only terms up to second order of magnitude, we transform the
equations (\ref{aa}) and (\ref{dis3}) to the following system of two
equations:
\begin{equation*}
i\hbar \frac{\partial \psi \left( \xi ,t\right) }{\partial t}=\left[ \lambda
-\frac{\beta D}{R}\rho \left( \xi ,t\right) -\frac{\beta \gamma D}{2R^{2}}%
\rho ^{2}\left( \xi ,t\right) \right] \psi \left( \xi ,t\right) -J\frac{%
\partial ^{2}\psi \left( \xi ,t\right) }{\partial \xi ^{2}}  \label{eqmot2}
\end{equation*}%
and
\begin{eqnarray}
&&\frac{\partial ^{2}\rho \left( \xi ,t\right) }{\partial t^{2}}-v_{a}^{2}%
\frac{\partial ^{2}\rho }{\partial \xi ^{2}}+\frac{\beta D}{RM}\frac{%
\partial ^{2}}{\partial \xi ^{2}}\left\vert \psi \left( \xi ,t\right)
\right\vert ^{2}+  \notag \\
&&+\frac{\beta \gamma D}{M}\left( \rho \frac{\partial ^{2}\left\vert \psi
\right\vert ^{2}}{\partial \xi ^{2}}+2\frac{\partial \rho }{\partial \xi }%
\frac{\partial \left\vert \psi \right\vert ^{2}}{\partial \xi }+\frac{%
\partial ^{2}\rho }{\partial \xi ^{2}}\left\vert \psi \right\vert
^{2}\right) =0 ,  \label{eqmot3}
\end{eqnarray}%
with
\begin{eqnarray}
\lambda &\equiv &\varepsilon +T+U-D-2J  \notag \\
T+U &=&\frac{M}{2}\left\{ \int\limits_{-\infty }^{\infty }d\xi \left(
\int\limits_{-\infty }^{\infty }d\eta \frac{\partial }{\partial t}\left\vert
\psi \left( \eta ,t\right) \right\vert ^{2}\right) ^{2}\right.  \notag \\
&&+\left. v_{a}^{2}\int\limits_{-\infty }^{\infty }d\xi \left\vert \psi
\left( \xi ,t\right) \right\vert ^{4}\right\} ,  \label{tu}
\end{eqnarray}%
and $v_{a}=\left( \omega /M\right) ^{\frac{1}{2}},$ where $v_{a}R=V_{a}$ is
the acoustic longitudinal velocity in the chain.

We will look for traveling solutions moving along the chain with some
velocity $V=vR.$ In this case, we have the following transformation
\begin{equation}
\rho \left( \xi ,t\right) =\rho \left( \xi -vt\right) ,\qquad \psi \left(
\xi ,t\right) =\Phi \left( \xi -vt\right) \exp \left\{ i\theta \left( \xi
,t\right) \right\} .  \label{travelling}
\end{equation}%
Replacing the equation (\ref{travelling}) in equation (\ref{eqmot3}) and
after integrating, we obtain
\begin{equation}
\rho =\frac{\beta D}{RM}\left( \frac{\left\vert \psi \right\vert ^{2}}{%
G-\gamma \frac{\beta D}{MR^{2}}\left\vert \psi \right\vert ^{2}}\right) ,
\label{ro2}
\end{equation}%
with $G=\left( v_{a}^{2}-v^{2}\right) $.

Finally, we substitute the equation (\ref{ro2}) into the equation (\ref%
{eqmot2}) and obtain the nonlinear equation
\begin{equation*}
i\hbar \frac{\partial \psi }{\partial t}+J\frac{\partial ^{2}\psi }{\partial
\xi ^{2}}-\left[ \lambda -\frac{k_{1}\left\vert \psi \right\vert ^{2}}{%
G-k_{3}\left\vert \psi \right\vert ^{2}}-\frac{k_{2}\left\vert \psi
\right\vert ^{4}}{\left( G-k_{3}\left\vert \psi \right\vert ^{2}\right) ^{2}}%
\right] \psi =0,  \label{saturation}
\end{equation*}%
with the values $k_{1}\equiv \frac{\beta ^{2}D^{2}}{R^{2}M},$ $k_{2}\equiv
\gamma \frac{\beta ^{3}D^{3}}{2R^{4}M^{2}},$ $k_{3}\equiv \frac{\beta \gamma
D}{MR^{2}}$. Rewriting these parameters in terms of the exciton-phonon
coupling constant $\chi =\frac{\beta D}{R}$, we have $k_{1}\equiv \frac{\chi
^{2}}{M},$ $k_{2}\equiv \chi ^{3}\frac{\gamma }{2RM^{2}}$. Equation (\ref%
{saturation}) is the well known NSE with saturable nonlinearity. This
equation arose earlier in various branches of physics, particularly in
nonlinear optics and simulated saturation (decrease) effects of the
nonlinear response of a medium in large electromagnetic fields \cite{makha2}

Let us further simplify equation (\ref{saturation}). If we take into account
in this equation, the nonlinearity not higher than $0(b\left\vert \psi
\right\vert ^{2}),$ and $k_{2}>>k_{1}k_{3}$, we obtain for the distribution
of excitations the Cubic - Quintic Nonlinear Schrodinger Equation(CQNSE)
\begin{equation}
i\hbar \frac{\partial \psi }{\partial t}+J\frac{\partial ^{2}\psi }{\partial
\xi ^{2}}-\lambda \psi +\frac{k_{1}}{G}\left\vert \psi \right\vert ^{2}\psi +%
\frac{k_{2}}{G^{2}}\left\vert \psi \right\vert ^{4}\psi =0.  \label{finalCQ}
\end{equation}

As known, nonlinear equations similar to equation (\ref{finalCQ}) possess
interesting structures when the attractive and repulsive terms could
compensate each other. So, in the next section we report some solutions that
appear as natural excitations along the molecular protein chain.

\section{Soliton Structures}

For solving the equation of motion presented in the previous section we have
to consider physically boundary conditions. Since we are interested on the
fact that the displacements of the perturbed units could only take the local
character, it is proposed that at long distances from the occurring
perturbations, displacements are very weak and practically the distribution
of excitations at long distance vanishes \textit{i.e} at " infinity" it is
zero. The second boundary condition is considered when the displacements
take constant values at infinity. These restrictions of our chain at
"infinities" could be fixed for the time evolution of the perturbations
along the chain.

\subsection{Trivial boundary condition.}

For simplicity, let us analyze the case when $k_{2}<0$ and $k_{1}>0.$\ If
this is done, the last term in (\ref{finalCQ}) represents the repulsive part
of the nonlinearity and the 4th term the attractive one. Further, if we make
the variable transformation $\tau =\frac{t}{\hbar }$ , $z=\sqrt{\frac{\kappa
}{J}}\xi ,\;\mu =\frac{\lambda GM}{\chi ^{2}},\;\;\nu =-\chi \frac{\gamma }{%
2MRG},$ we finally obtain:

\begin{equation}
i\frac{\partial \psi }{\partial \tau }+\frac{\partial ^{2}\psi }{\partial
z^{2}}-\mu \psi +\left( \left\vert \psi \right\vert ^{2}-\nu \left\vert \psi
\right\vert ^{4}\right) \psi =0.  \label{finalCQ2}
\end{equation}

The CQNSE (\ref{finalCQ2}) was studied from various points of view, here we
follow the results and conclusion obtained in the works \cite{makha},\cite%
{peli2} and \cite{me}.

The corresponding solution of the CQNSE with trivial boundary condition
\begin{equation*}
\psi \rightarrow 0\text{\ \ \ for \ \ }x\rightarrow \pm \infty ,
\end{equation*}%
is the so called drop-type soliton that is not a topological soliton because
the vacuum also has the same asymptotic value. This implies that for the
equation (\ref{finalCQ2}), we have the static non-topological soliton \cite%
{makha}. The non topological soliton solutions are those of which boundary
conditions at infinite are the same that vacuum state. However, topological
solitons have boundary conditions different from the vacuum. This means in
particular, that states of degenerated vacua might exist. It is to say that
soliton "will be moored" by its boundary conditions. An example of
topological soliton solution is a step or kink. For our case we have the
drop soliton

\begin{equation}
\psi =e^{i\theta _{0}}\sqrt{-4\alpha }\left( 1+\sqrt{1+\frac{16\alpha }{3}}%
\cosh \left( \sqrt{-\alpha }\left( x-x_{0}\right) \right) \right) ^{-\frac{1%
}{2}},  \label{drop}
\end{equation}%
with
\begin{equation}
\alpha =-\mu \nu =-\frac{\lambda \gamma }{2\beta D}.  \label{alfa}
\end{equation}%
The traveling soliton should be obtained using the Galileo transformation in
the same form what we define a traveling wave in mechanics

\begin{eqnarray*}
\theta _{0} &\rightarrow &\frac{V}{2}x-\frac{V^{2}}{4}t+\theta _{0}, \\
\cosh \left( \sqrt{-\alpha }\left( x-x_{0}\right) \right) &\rightarrow
&\cosh \left( \sqrt{-\alpha }\left( x-Vt-x_{0}\right) \right) .
\end{eqnarray*}

The soliton solution (\ref{drop}) has the normalized motion integral named
the \textquotedblleft number of particles\textquotedblright\ $\int
dx\left\vert \psi \right\vert ^{2}=1,$ calculating this integral we find
\begin{equation*}
\frac{16\alpha }{3}=\frac{1}{\cosh [1/\sqrt{3}]}-1.
\end{equation*}%
The approximate value for the parameter $\alpha \approx -0.051$. Replacing
this value in the relation (\ref{alfa}) we obtain the restrictions of the
main parameters $\frac{\lambda \gamma }{\beta D}=0.102$. The quintic part of
the nonlinear equation produces the effect of counterbalance the attractive
forces between \textquotedblleft two particles \textquotedblright\ in the
mechanical analogy method represented by the cubic nonlinearity.

If we replace (\ref{drop}) in (\ref{ro2}) we have the distribution of
changes in the relative distance between molecules.

\begin{equation}
\rho (\xi ,t)=\left( \frac{1-\sigma }{\sigma \gamma R}\right) \frac{1}{%
1+\eta \cosh \left[ \sqrt{-\alpha }\left( \xi -\xi _{0}\right) \right] },
\label{dis1}
\end{equation}%
with $\sigma =1+\frac{4\alpha B\gamma D}{MG}=1-\frac{2\lambda \gamma ^{2}}{MG%
},\;\eta =\frac{\sqrt{1+\frac{16\alpha }{3}}}{\sigma }.\;$

\begin{figure}[tbp]
\includegraphics[width=8cm]{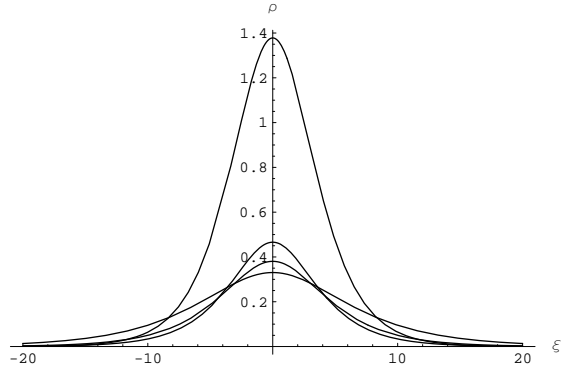}
\caption{Displacement of the molecules represented by the equation ( \protect
\ref{dis1}) around the equilibrium position for $\protect\lambda =2.5,
\protect\sigma=0.1,\protect\gamma=2.3,R=1.2$}
\label{fig1}
\end{figure}

Some numerical representations of this solutions with different values of
the main parameters are represented in the figure \ref{fig1}.

In the particular case when the parameters satisfy the relation
\begin{equation*}
\frac{8}{3\beta D}=\frac{\gamma }{MG}\left( 1-\frac{\lambda \gamma ^{2}}{MG}%
\right) ,
\end{equation*}%
we have the solution
\begin{equation}
\rho (\xi ,t)=\frac{\lambda \gamma }{R\left( 1-\frac{2\lambda \gamma ^{2}}{MG%
}\right) }\frac{1}{\cosh ^{2}\left[ \frac{1}{2}\sqrt{-\alpha }\left( \xi
-\xi _{0}\right) \right] },  \label{diss2}
\end{equation}%
which is the new distribution of changes in the relative distance between
particles.

From (\ref{dis1},\ref{diss2}) the maximum deviations $d_{1}$ and $d_{2}$ are
\begin{eqnarray*}
d_{1} &=&\frac{2\lambda \gamma }{MGR\left( 1-\frac{2\lambda \gamma ^{2}}{MG}+%
\sqrt{1-\frac{8\lambda \gamma }{3\beta D}}\right) }, \\
d_{2} &=&\frac{\lambda \gamma }{R\left( 1-\frac{2\lambda \gamma ^{2}}{MG}%
\right) }.
\end{eqnarray*}%
From these equation we can see that the displacement due to the appearance
of solitons corresponding to Eq. (\ref{dis1}) is greater compared with the
similar equation (\ref{diss2}). This means that the strong "damage" will be
caused by the soliton represented by the equation (\ref{diss2}). We can see,
that the presence of solitons leads to the pronounced deviation of the
peptide groups from their equilibrium position and the second one should
produce a breakage of the chain system when $\frac{2\lambda \gamma ^{2}}{MG}%
=1.$

Using the expression (\ref{tu}), we can obtain the values of the total
energy of the peptide group displacement as follows:

\begin{equation*}
E_{1}=\frac{16}{\left( \sqrt{\frac{16}{3}\alpha }\right) ^{3}}\sqrt{\frac{%
-\alpha J}{k}}\left( \frac{J\nu }{\hbar k}+4\alpha _{0}^{2}\right) \arctan
\left( \frac{\sqrt{1+\frac{16}{3}\alpha }}{\left( \sqrt{\frac{16}{3}\alpha }%
\right) ^{3}}\right) .
\end{equation*}

\subsection{Condensate Boundary conditions}

We can suggest also, that there are very specific restrictions that could
cause soliton excitations to appear along the protein chain. Besides the
natural or well known bell soliton excitation it is also very possible the
appearance of other types of solutions. For example, there could surge
topological or non topological solitons because the CQNSE supports them. It
is well known that the equation (\ref{finalCQ}) supports kinks and bubble
type of solitons. For this case, we will use the well reported results
obtained by many authors specifically we mention \cite{makha}.

Let us see the case of regular solutions of the CQNSE (\ref{finalCQ}) in 1+1
dimensions of the space-time. We rewrite this equation in a slightly
different form by using the ground states and putting them in the equations.
In order to visualize the ground states, we will use the following form of
the CQNSE:

\begin{equation}
i\varphi _{\tau }+\varphi _{\varsigma \varsigma }-\left( 3\left\vert \varphi
\right\vert ^{2}-\left( 2A+\sigma _{0}\right) \right) \left( \left\vert
\varphi \right\vert ^{2}-\sigma _{0}\right) \varphi =0.  \label{eqfi}
\end{equation}

This version permits us to find the soliton solutions in explicit form. It
can be demonstrated that the eq. (\ref{eqfi}) could be generated from the
relation (\ref{finalCQ2}) with the help of the following scale
transformations

\begin{equation}
\varphi \left( \zeta ,\tau \right) =\sqrt{\frac{3}{2\nu \left( A+2\rho
_{0}\right) }}\psi \left( x,t\right) ,  \label{scali}
\end{equation}

\begin{equation*}
\tau =\left( \frac{9}{8}\right) \frac{1}{\nu }\left( A+2\rho _{0}\right)
^{-2}t,\ \ \ \ \zeta =\frac{3}{2}\frac{1}{\sqrt{2\nu }}\left( A+2\rho
_{0}\right) ^{-1}x.
\end{equation*}%
Making some algebra, parameters $A$\ and $\sigma _{0}$ are related each
other by the relation
\begin{equation}
\frac{A}{\sigma _{0}}=-2+\frac{3}{4}\frac{1}{\mu \nu }\left( 1+\sqrt{1-4\mu
\nu }\right) .  \label{param}
\end{equation}

Without loss of generality it is possible to fix the value of $\sigma _{o}=1$%
, because properties of the solutions depend easily on the parametric
relation $(A/\sigma _{o})$. Here the parameter $A$ can be both positive or
negative based on the physical requirement we could impose on the system.
Further, the variables $\zeta $ and $\tau $ will be treated as if $x$ and $t$
were the customary variables.

\subsubsection{Bubble solitons.}

\begin{figure}[tbp]
\includegraphics[width=8cm]{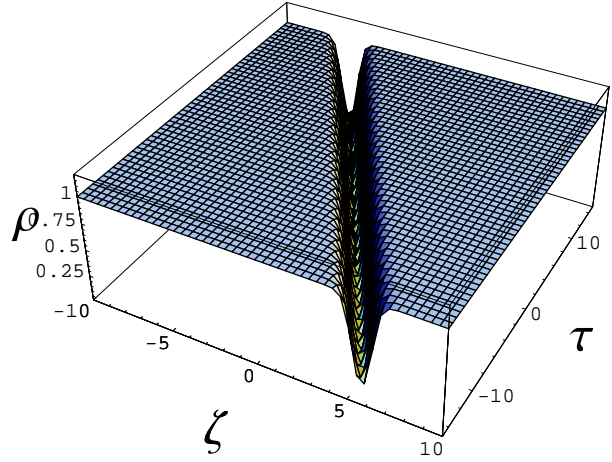}
\caption{This displacement is similar to the well known "classical bubble"
solution in field theory, but for our case it represents the agglomeration
of units in the molecular chain and for this reason it is named "crowdon" (
\protect\ref{dispo})}
\label{fig2}
\end{figure}

For obtaining gray or bubble solitons when the degenerated vacuum is not
absolute, it is imposed the boundary condition
\begin{equation*}
x\rightarrow \pm \infty \text{ \qquad and \qquad }\varphi \rightarrow \kappa
_{3}.
\end{equation*}%
with $\kappa _{3}=\sqrt{\frac{2A+1}{3}}$ . For this case, the solution of
equation (\ref{eqfi})takes the form

\begin{equation}
\varphi _{b}=\frac{\sqrt{4-A}}{\sqrt{3}}\frac{e^{\left( i\theta \right)
}\cosh \left[ 2\sqrt{a}(y-y_{0})\right] }{\sqrt{1+\frac{4-A}{2A+1}\sinh ^{2}%
\left[ 2\sqrt{a}(y-y_{0})\right] }},  \label{bubo}
\end{equation}%
with $y=\zeta -v\tau $ and
\begin{equation*}
a=\frac{1}{4}\left( v_{s}^{2}-v^{2}\right) =\frac{1}{4}\left( \frac{4}{3}%
(A-1)(1+2A)-v\right) ,
\end{equation*}%
This bubble is a nontopological solution, and their topological charge is
equal to zero, $Q_{b}=0$. Here the parameter $A$ satisfies $1<A<4$. then for
the displacement $\rho $ according to the equations (\ref{ro2}) and (\ref%
{bubo}) we obtain the expression
\begin{equation}
\rho =\frac{3R}{2\gamma }\left[ \frac{\cosh ^{2}\left[ 2\sqrt{a}(y-y_{0}%
\right] }{\alpha _{1}+\alpha _{2}\sinh ^{2}\left[ 2\sqrt{a}(y-y_{0}\right] }%
\right]  \label{dispo}
\end{equation}%
being
\begin{equation*}
\alpha _{1}=\alpha _{2}+9G\frac{(A-1)}{(2A+1)(4-A)},
\end{equation*}%
and
\begin{equation*}
\alpha _{1}=\frac{3G}{4-A}-\kappa ,\ \ \ \alpha _{2}=\frac{3G}{2A+1}-\kappa ,
\end{equation*}%
while
\begin{equation}
\kappa =\frac{m\gamma \beta D}{M}\ , \ m=\frac{2\nu (A+2)}{3}.  \label{kappa}
\end{equation}%
In the case when the $\alpha _{1,2}$ are both positive, it is easy to see
that $\alpha _{1}>\alpha _{2}$ since for this case $4>A>1$. The displacement
corresponding to the bubble like soliton excitation $\varphi _{b}$ is a
typical gray soliton and can be depicted in Fig \ref{fig2}. Apparently this
type of solution is similar to others obtained for nonlinear classical
models. But in contrast to the well known feature, in our case we have not a
bubble displacement, instead we have an agglomeration of molecules that
travels along the chain i.e. we have here typical crowdon solution forming
the agglomeration of molecules. This agglomeration is traveling along the
chain like an accumulation of molecules conserving velocity and profile.

\begin{figure}[tbp]
\includegraphics[width=8cm]{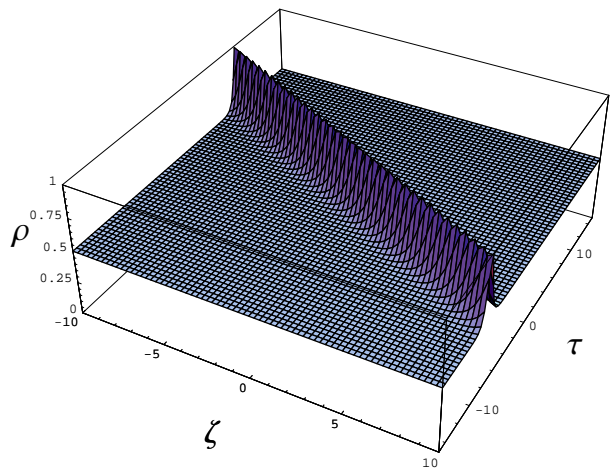}
\caption{Typical soliton like perturbation on the background that represents
the displacements. The interpretation that could be given to this solution
is that in a real manner represents a bubble or the region inside of which
the displacements are increasing and this effect is contrary to the
agglomeration. }
\label{fig3}
\end{figure}

This result is linked to solutions that are moving slowly with less velocity
in comparison with the sound velocity, \textit{i.e.} when $v_{s}^{2}-v^{2}>0$%
. But when the opposite occurs, \textit{i.e.} when the soliton velocity is
greater than the sound velocity we have soliton solution on the step or on
the background. This type of solution is presented in the Fig. \ref{fig3}.

So, the sign of $G$ determines the type of soliton solution for the
displacement that corresponds to the distribution of excitations. When $%
1<A<4,$ bubble solitons for the excitations appear corresponding to
displacements of both types: popular gray solitons and normal "bell" soliton
on the background. In the first case for displacements we have obtained the
crowdon solution. In the last case, this displacement shows a huge
separation between neighborhoods in the molecular chain. This solution
travels along the protein with velocity $v$ greater than the acoustic
velocity. Of course, we have here a little difference of popular bubbles of
other nonlinear equations. As it is well known, the classical bubble is
characterized by the rarefaction inside crucial region, but in this case,
the feature of our solution is contrary to this best known property of
classical bubbles. Indeed, the displacements outside the perturbed region
have constant values, while inside of this region the separation between
molecules increase as shown in Fig. \ref{fig3}. When this displacement is
strong enough, one could observe the breaking of the protein chain and the
solution could be transformed to a peak or cusp like non-classical solution.

\subsubsection{Crowdon solutions}

The "space" of our model is a line with two points as boundaries at the
infinite. The existence of topological soliton named kink, is due to the
properties of the mapping between the degenerated minimums of the potential,
with the space, saying better, with its boundaries, that in this case it is
a discrete set.

Kink solitons appear when the potential supports degenerated absolute vacua
in $\varphi =\kappa _{3}=\frac{2A+1}{3}$, when $A>4$. So, the boundary
conditions for this case, are
\begin{equation*}
\varphi \left( \zeta =-\infty \right) =-\kappa _{3},\;\varphi \left( \zeta
=\infty \right) =\kappa _{3}.
\end{equation*}

The kink solution has the following form
\begin{equation}
\varphi _{k}=\frac{\sqrt{A-4}}{\sqrt{3}}\frac{e^{i\theta }\sinh [2\sqrt{a}%
(y-y_{0})]}{\sqrt{1+\frac{A-4}{2A+1}\cosh ^{2}[2\sqrt{a}(y-y_{0})]}}.
\label{kinki}
\end{equation}
For antinkink, as usual, we can take the inverse signs, so when $%
x\rightarrow +\infty \,$, the field will approach the value $-\kappa _{3},$.
Calculating the topological charge of this solution we obtain $Q_{k}=1$. For
the antikink, we have $Q_{ant}=-1$. In this case, the regular solutions with
finite energy are divided in 4 topological sectors. The sectors with finite
energy can be characterized by means of the following pairs of indices $%
(-\kappa _{3},\kappa _{3}),(\kappa _{3},-\kappa _{3}),(-\kappa _{3},-\kappa
_{3})$ and $(\kappa _{3},\kappa _{3})$ that correspond to the values of the
field at infinities \textit{i.e.} for $\varphi \left( x=-\infty \right)
,\;\;\varphi \left( x=+\infty \right) $. The kink, antikink and trivial
solutions $\varphi \left( x\right) =\pm \kappa _{3}$ are the members of the $%
4$ sectors.

Using the equations (\ref{kinki}) and (\ref{ro2}) For the displacement $\rho
(\zeta ,\tau )$ we have the following expression
\begin{equation}
\rho (\zeta ,\tau )=\frac{m\beta D}{RM}\left( \frac{\sinh ^{2}\left[ 2\sqrt{a%
}(y-y_{0})\right] }{\alpha _{3}+\alpha _{4}\cosh ^{2}\left[ 2\sqrt{a}%
(y-y_{0})\right] }\right) ,  \label{diskink}
\end{equation}%
with
\begin{equation*}
\alpha _{3}=\frac{3G}{A-4}+\kappa ,\ \ \ \ \alpha _{4}=\frac{3G}{2A+1}%
-\kappa ,
\end{equation*}%
and $\kappa $ is defined in equation (\ref{kappa}).

\begin{figure}[tbp]
\includegraphics[width=8cm]{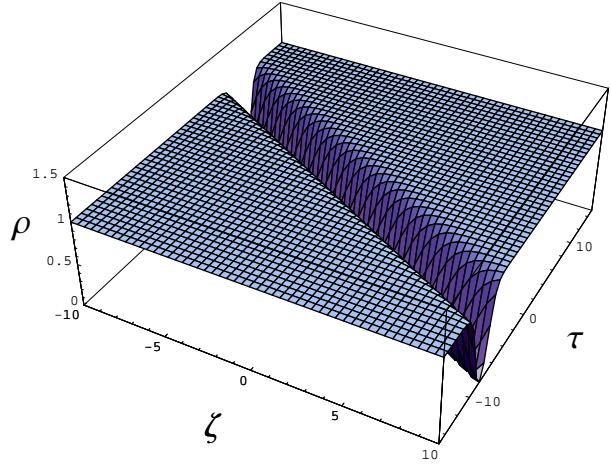}
\caption{The evolution of the displacements shows typical characteristic of
"crowd" soliton solution. For a qualitative representation of the solution,
we take the values $\protect\alpha_4 = 1, \protect\alpha_3 = 3, v = -1.2$.
Apparently, this picture represents "classical bubble" solitons but inside
the perturbed region the displacements reduce their values forming the
crowdons (\protect\ref{diskink}).}
\label{fig4}
\end{figure}

Considering positive values for $\alpha _{i}$ when the boundaries values of
displacements are fixed, there is a possibility of emergence of crowdons
that will propagate along the chain. The displacements show the typical
picture of "classical" bubble solitons. This bubble in the molecular chains
context is not a dip in the background. On the contrary, this solutions
represent the wave agglomeration of molecular units traveling along the
chain. So, along the molecular chain the distribution of excitations evolves
like a kink soliton while its corresponding displacement evolves like
typical crowd like soliton solutions, see Fig. \ref{fig4}.

\subsection{Stability and Self-localizations}

Next, we can use some results concerning the possibility of stabilizations
of those types of solitons we have found in the section \textbf{A} and
\textbf{B} for those distribution of excitations with non- topological
charges. We will follow the findings of the paper \cite{bernal}

As it is well known, our system is closely related to the phi-six theory,
that shows several important behavior in nuclear hydrodynamics,
ferromagnetism, phase transitions and other branches of physics and natural
sciences \cite{makha}. The equation of motion that we will analyze in the
context of self localization is the equation (\ref{finalCQ2}) obtained above.

The main ideas of the approach developed in \cite{makha} and \cite{bernal}
could be summarized as follows: This method in some sense is similar to
Lyapunov functional approach for analyzing stability problems. For the
general multidimensional case of spherical or cylindrical symmetry the
dynamical system associated with the solitons NSE is:\medskip

\begin{equation}
i\psi _{t}+\psi _{rr}+\frac{\mathcal{D}-1}{r}\psi _{r}-\frac{dU(\Phi )}{%
d\Phi }\psi =0,  \label{uno}
\end{equation}%
where $\Phi \equiv $ $\psi ^{\ast }\psi $ , the dimension of the space is $%
\mathcal{D}$ and with suitable change of parameters we use the potential
part of the energy as \cite{makha,bernal}

\begin{equation*}
U\left( \Phi \right) =U(\Phi )=\Phi +2A\Phi -\left( 2+A\right) \Phi
^{2}+\Phi ^{3},
\end{equation*}%
with
\begin{equation*}
A=-2-\frac{3}{4\alpha }(1+\sqrt{1+4\alpha }),
\end{equation*}%
and $\alpha $ is defined by the equation (\ref{alfa}). For our case, a
molecular system, the equation (\ref{uno}) is the same equation (\ref%
{finalCQ2}) with $\mathcal{D}=1$ \cite{me}. The constant solutions of the
equation (\ref{finalCQ}) at $A=-\frac{1}{2}$ undergoes a supercritical
bifurcation.

For analyzing the self-localizations of soliton structures we will evaluate
the behavior of the following functional $B(t)$ \cite{makha,bernal}

\begin{equation}
B(t)\equiv \pi 2^{\mathcal{D}-1}\int \Psi ^{\ast }\Psi r^{D+1}dr=\langle
r^{2}\rangle N\geq 0,  \label{bt}
\end{equation}%
where $N$ denotes the \textquotedblright number of
particles\textquotedblright\ integral of motions. The magnitude $B(t)$
represents in some sense the width of the soliton solution obtained in this
model. As it can be seen from the equation (\ref{bt}) the functional $B(t)$
is proportional to the expectation value of the square of the position. From
here this magnitude is proportional to the width of the soliton solution.
Let us evaluate the second derivative of this magnitude and we obtain
\begin{eqnarray}
\frac{d^{2}B(t)}{dt^{2}} &=&  \notag \\
&&2^{\mathcal{D}-1}\pi \left\{ 8\int \left\vert \psi _{r}\right\vert ^{2}dr+4%
\mathcal{D}\left[ \int \left( \frac{dU}{d\Phi }\Phi -U\right) dr\right]
\right\} .  \label{b}
\end{eqnarray}

As we are concerned with solitons in one dimensional case for the molecular\
system, finally making $\mathcal{D}=1,$ and evaluating the conditions under
self-localization of soliton structures we finally obtain that following
conditions \cite{bernal}

\begin{equation}
\text{If }\,\mathcal{D}=1,\;\;\;\frac{\delta \ddot{B}}{\delta \mathfrak{V}}%
<0,\,\;\;\;\text{for \ \ \ }A<2,  \label{self}
\end{equation}%
where $\mathfrak{V}$ denotes the \textquotedblright
volume\textquotedblright\ of the soliton structure. Now let us calculate the
possible values of the parameters for getting self localizations of
solitons. We make $\alpha =-\varrho $ and for satisfying the condition (\ref%
{self}) we obtain the inequality
\begin{equation}
\frac{1}{\varrho }\left( 1+\sqrt{1-4\varrho }\right) <\frac{16}{3}
\label{selfloca}
\end{equation}%
with $\varrho =\frac{\lambda \gamma }{2\beta D}.$ As we can see from the
relation (\ref{selfloca}) for all values of $\varrho <0,$ the
self-localization is possible \textit{i.e.} when the parameters satisfy this
relation $\frac{\lambda \gamma }{2\beta D}<0$. In the case of positive
region of the axis $\varrho$, we can conjecture that self-localizations
would occur when
\begin{equation*}
0.2345\leq \frac{\lambda \gamma }{2\beta D}\leq 0.25  \label{conje}
\end{equation*}%
That means, we should try to find stable soliton like solutions within this
sector that could determine the possible values of the relevant parameters
in real experiments.

\section{Conclusions}

We have taken into consideration in some sense an improved Davydov`s model,
and obtained a nonlinear evolution equation for the displacement that in
mutual coordination with the excitation generate nontopological and
topological solitons.

When the molecular system evolves, it is of course interesting to know which
types of structures it could support during its evolution when its dynamic
is modeled by the cubic quintic nonlinear Schrodinger equation. When the
boundary condition is of the trivial one, we obtained a typical well known
"bell" soliton that represents the separation of the molecular units inside
the perturbations. Let us make some comments about the physical meaning of
the solution obtained when we considered the condensate type of boundary
conditions. In this specific case, the molecular model subjected to both
"condensate boundary conditions", in both directions, the displacements $%
\rho $ acquires a constant value. This means that due to multiple  internal
or external factors that affects the molecular system, the displacement far
from the perturbed region acquires some certain values that in some sense
could be considered as established during the process of the existence. In
the case, when bubble like solutions appear as natural excitations,
displacements around the central part of the perturbed region could be of
two types: the first one corresponds to the crowd solutions, that means we
have in some sense an agglomeration of units that compound the lattice i.e.
a "crowdon". This agglomeration travels along the chain with velocity $v_{s}$
less than that of the sound. The second solution for displacements is the
"soliton on the step" and travels faster than sound. It represents the
increasing of distance between elemental units inside the perturbed region
of the molecular chain. For the case, when the excitations reveal the
typical profile of kink solutions, as we can see from equations (\ref%
{diskink}), the displacements apparently evolve like "classical bubble"
solitons but they represent the crowdons again, and represent the
agglomeration of units along the molecular chain. According to the stability
analysis, we also conjecture that it is possible to obtain stable solitons
inside the narrow segment in the parametric domain determined by the
equation (\ref{conje}). Finally, it should be interesting to check in more
realistic models, the behaviors of these structures after the inclusion of ,
for example, dissipation term to describe the effects from water molecules
surrounding the molecular system.

\begin{acknowledgments}
We are grateful to Prof. Makhankov for valuable discussions. This work was
partially supported by CONACYT 61401 Research Project and  the PROMEP grant
for supporting research groups and carried out in part during the MAG visit
to San Diego State University. RG-S acknowledges partial support from COFAA
and to the grants EDI and SIP 20080759.
\end{acknowledgments}

\end{document}